\documentclass[12pt]{article}
\begin{document}
\title{Computers Should Be Uniters Not Dividers:\\A Vision of
Computer-Enhanced\\Happy Future}
\author{Alexander Titovets$^1$, Philip Mills, and Vladik Kreinovich$^2$\\
$^1$El Paso, Texas, USA, 2titovetsart@gmail.com\\
http://www.titovetsart.com\\
$^2$University of Texas at El Paso, vladik@utep.edu}
\date{}
\maketitle

\begin{abstract}
This manifesto provides a vision of how computers can be used to
bring people together, to enhance people's use of their natural
creativity, and thus, make them happier.
\end{abstract}

\noindent{\bf Computers could be uniters but they are dividers.}
Computer networks connect people from all over the world. This
should make people feel closer to each other -- but instead, it
divides.

In the past, when there were only a few TV channels, a few famous
books, a few new movies, people had a common ground, something to
talk about and discuss. Not anymore. In the US, Democrats watch
their own news, have their own chat groups, Republicans have their
own, Democrats and Republicans rarely engage in discussions.
People who collect stamps chat together, people who love ballet
chat together, different interest groups rarely mingle.

A scientist goes to a conference abroad. He or she does not need
to (and probably does not want to) get local news: if there are no
familiar US channels on the hotel TV, there is always a computer
access.
\medskip

\noindent{\bf But maybe divisions are good?} Maybe nothing is
wrong with this division, maybe people are happy this way?

Not really. Many people -- especially after having achieved a
certain level of success -- are bored; see, e.g., \cite{Kraut
1998,Wellman 2001}. Talking only to people who share your views is
as exciting as talking to yourself. And talking to people with
different sets of interests is also boring: any professional knows
what happens when you start describing your exciting professional
challenges to a stranger at a party :-(

So what can we do about it?
\medskip

\noindent{\bf What do people want?} Since the problem is that
people are bored, people are not very happy, a natural answer is
to ask people what they want.

Many people readily provide an answer: they want to contribute,
not just contribute by writing checks (this is important but not
that exciting), not just contribute by using themselves as
low-level menial workers as when building houses for Habitat for
Humanity (rewarding, important, but not that exciting). What many
people want is to find some activity where they can use not only
their checkbook and their hands, but their {\it creativity}; see, e.g.,
\cite{Barron 2012,Carr 2011,Carson 2010,Csikszentmihalyi 1990,Layard
2011,Norris 2013,Zeki 2011}.
People want to make contributions which are distinctly, creatively
theirs.

Many retired people take art classes, take classes in other areas,
and become much happier \cite{Barron 2012,Carr 2011,Carson 2010,Csikszentmihalyi 1990,Layard
2011,Norris 2013,Zeki 2011}.
\medskip

\noindent{\bf People want to be creative, but can they?} Maybe
this is an illusion? Maybe in every generation, there are a few
creative geniuses, but the rest of us have no talent for
creativity?

Luckily, this is not true. For example, when retired people go
into art, they often generate interesting creative objects, so
there clearly is a creativity spark in most of us -- and maybe in
all of us.
\medskip

\noindent{\bf So maybe art classes are a solution?} Alas, no.
There are not too many artists in each community, and even fewer
artists who can teach. Same goes for other intellectual endeavors.

And also, a professional teacher can appreciate the creativity of
a person, but for others to enjoy the result of this creativity,
this result needs to be professionally improved: amateur stories
must be professionally edited, amateur movies must be re-made
professionally -- and there is no time for that on a mass scale.
\medskip

\noindent{\bf Technology can help.} How can technology help? Let
us start with an analogy. People have always loved to hear good
signing. In the ancient times, an emperor could afford to
entertain himself and his guests with the world's best singing.
Now, with TV and computers, operas from the Metropolitan Opera and
other leading opera houses are streamlined all over the world.
With recording devices, we can enjoy these operas at any time.
\medskip

\noindent{\bf So what shall we do?} People want to contribute, to
improve the life on Earth, to contribute creatively. We need to
design an infrastructure for enabling them to do it.

The closest we are right now to such an infrastructure is
Wikipedia (and the web in general). It is used by everyone, and,
in principle, everyone can contribute. However, this is still not
easy -- and besides, Wikipedia just accumulates knowledge, it does
not produce anything new. It is helpful, useful -- but it is still
not that creative and, honestly, often rather boring. And
Wikipedia only helps with knowledge, it is of not much use to
whose who are more artistically inclined.
\medskip

\noindent{\bf So what shall we do?} How are new things created? It
is rare that an Einstein just sits alone and come up with all the
new ideas. Usually, an Einstein talks to other Einsteins, reads
papers and books by others -- and it is this interaction that
serves as a breeding ground for creativity. So, if we want
creativity, this is what we need to emulate: discussions with
genuises and between genuises.

To some extend, an interaction with a genius is what we do when we
read a Dostoevsky novel or Einstein's paper or listen to Bach's
music. But this is a limited interaction. We cannot ask them a
question, we cannot change what they have written -- and although
sometimes we wish to hear a dialogue between Jesus Christ and
Buddha, they never met -- so this wish cannot be fulfilled. Or can
it?
\medskip

\noindent{\bf This is what we need to do.} Einstein dies in 1955,
but we know a lot about him -- from his writings, from the
memories of people who have known him. We can often reasonably
predict how he would react to different events in the world, to
different opinions of others -- and if we do not know, we often
speculate. What Would Jesus Do, What Would Buddha Do -- these
slogans have become, for many, a way to live.

So this is what we need to emulate. We need to create a virtual
universe in which there will be avatars of great geniuses of the
past -- computer programs that try their best to simulate the
geniuses' ways to thinking; see, e.g., \cite{Atkins 2011,Bell
2001}. And if we are not sure, if we have several hypotheses about
what an ancient genius really thought -- well, there is nothing
wrong with designing several different computer versions of that
genius. And maybe we should make several versions -- for example,
instead of a single computer model of Picasso, maybe a good idea
is to develop several Picasso-emulating models corresponding to
different stages in his life and in his art?

Something like this is done in computer games -- except that we do
not just want to senselessly shoot Nazis or zombies, we want to
make it creative. There are already avatars of Einstein and
others; see, e.g., \cite{Case 2007,Johnson 2008,Tang 2008} -- but
these current avatars are still short of Einstein's creativity.

The future creative avatar programs should be able to communicate
with each other -- and we will be able to be proud witnesses to
discussions involving Jesus, Buddha, and Einstein. We can witness
a two-way dialog between an early Picasso and a later Picasso,
with Michelangelo chiming in?

And where do common folks come in, with their sparks of
creativity? Well, these models of geniuses is what will help us to
learn how to unleash our internal creativity -- and what will help
us transform it into something that others can use and enjoy.
Einstein will edit texts containing our research ideas, Bach will
help us instrument our melodies -- and they may guide us into
using our talents for something that is most useful for humankind.

And since these are computer models, not real people, there is no
limitation on how many of us they can serve at the same time.
\medskip

\noindent{\bf How is this different from the original ideas of
Artificial Intelligence?} The idea of creating a computer-based
Einstein sounds suspiciously close to the naive over-optimistic
1950s ideas of creating an Artificial Intelligence -- a genius
computer that will solve all our problems.

But there is a difference. Yes, some problems will be solved by a
supercomputer, but most problems will be solved by us -- with the
supercomputer acting as helper, as an enhancer of our creativity.
\medskip

\noindent{\bf Let us hope.} In this brave new world, the virtual
reality will not be only an entertainment trick (as it is --
mostly -- now). It will be a medium connecting everyone on Earth
-- a medium in which avatars of geniuses of art and science will
teach us, help us, communicate with us, and we all together, real
and virtual, will help make this world a better and happier place.

Let us hope for this. And let us work to make this happen.

\end{document}